\documentstyle[psfig,amssymb]{mn}

\newif\ifAMStwofonts

\begin{document}

\title[
The analysis of integrated spectra from globular clusters
]
{
A Robust Method for the Analysis of Integrated Spectra from Globular Clusters 
using Lick Indices
}
\author[
Proctor, Forbes \& Beasley
]
{
Robert N. Proctor$^{1}$, Duncan A. Forbes$^{1}$ \& Michael A. Beasley$^{1,2}$\\ 
$^1$Centre for Astrophysics \& Supercomputing, Swinburne University,
Hawthorn VIC 3122, Australia\\
Email: rproctor@astro.swin.edu.au,
dforbes@astro.swin.edu.au\\
$^2$ Lick Observatory, University of California, Santa Cruz, 
CA 95064, USA\\
Email: mbeasley@ucolick.org}

\pagerange{\pageref{firstpage}--\pageref{lastpage}}
\def\LaTeX{L\kern-.36em\raise.3ex\hbox{a}\kern-.15em
    T\kern-.1667em\lower.7ex\hbox{E}\kern-.125emX}

\newtheorem{theorem}{Theorem}[section]

\label{firstpage}

\newpage

\maketitle

\begin{abstract}
We define a method for analysis of the  integrated spectra of extra-galactic 
globular clusters that provides more reliable measures of age, metallicity
and $\alpha$-element abundance ratio than have so far been achieved. The
method involves the simultaneous fitting of up to 25 Lick indices in a
$\chi^2$-fitting technique that maximises the use of the available data.
Here we compare 
three sets of single stellar population (SSP) models of Lick indices to the 
high signal-to-noise, integrated spectra of 20 Galactic globular clusters. 
The ages, [Fe/H] and $\alpha$-element abundance ratios derived from 
the SSP models are compared to the results of resolved stellar population
studies from the literature. We find good consistency with the
published values, with an agreement of better than 0.1~dex in all three 
derived parameters. The technique allows identification of abundance ratio
anomalies, such as the known nitrogen over-abundance in Galactic globular
clusters, and the presence of anomalous horizontal-branch morphologies. It
also minimises the impact on the derived parameters of imperfect calibration to the
Lick system, and reduction errors in general. The method defined in this work is
therefore robust with respect to many of the difficulties that plague the
application of SSP models in general, and is, consequently, well suited to the study 
of extra-galactic globular cluster systems.
\end{abstract}


\section{Introduction}
Even with the superior spatial resolution from space telescopes or adaptive 
optics, only for the very nearest galaxies will we be able to resolve 
individual stars. Thus the vast bulk of extra-galactic studies necessarily 
involve the analysis of integrated stellar populations. Models using Lick
indices (Worthey 1994; Trager et al. 1998) have been 
developed which allow derivation of key parameters such as age, metallicity and 
$\alpha$--element abundance ratio from the spectra of integrated stellar 
populations (e.g. Vazdekis 1999; Bruzual \& Charlot 2003; Proctor \& Sansom
2002; Thomas, Maraston \& Bender 2003). 
These models are best suited to single stellar populations 
(SSPs), but they are often applied to the composite stellar populations 
found in galaxies (e.g. Worthey, Faber \& Gonzalez 1992; Gorgas et al. 1997; 
Greggio 1997; Trager et al. 2000; Proctor \& Sansom 2002; Terlevich \&
Forbes 2002; Proctor et al. 2004)
              
Perhaps the best examples of single stellar populations that are easily
accessible to us are the Galactic globular clusters (GCs), for
which we have resolved colour-magnitude diagrams (e.g. Piotto et al 2002)
and/or high resolution spectra (e.g. Carney 1996). Indeed 
the SSP models are often built using such information from GCs. 
Thus, in order to have any confidence in the results of applying SSP models
to unresolved galaxies, it is necessary that the
SSPs correctly reproduce the properties of Galactic GCs. 
           
Here we apply three recent sets of SSP models to 20 Galactic globular
clusters with high signal-to-noise integrated spectra. 
We also compare the results to those from high resolution spectra and 
colour-magnitude diagrams (CMDs) from the literature. Previous studies of 
Galactic GCs using Lick indices (Puzia et al. 2002; Thomas
et al. 2003) have used the GC ages measured from
CMDs as an input to their interpretation of the indices.
However, this approach is unsuitable for the studies of
\emph{extra-galactic} GCs, as such \emph{a priori} knowledge of ages is 
unavailable. Instead, it is necessary to derive age, [Fe/H], and 
$\alpha$--element abundance ratios directly from the spectra.
In this way, we are able to test the SSP models and to place constraints on 
the accuracies that can currently be achieved for the ages and metallicities
of extra-galactic GCs using Lick indices.

\section{The Data and models}
\label{data}
We use data from two separate studies. The first is the 
data and spectra of 12 Galactic globular clusters (GCs) and the bulge of the
Milky Way from the study of Puzia et al. (2002; hereafter P02). 
These long--slit data were collected using 
the ESO 1.52 m telescope on La Silla with a dispersion of 1.89~$\AA$/pix 
and achieved a signal--to--noise in excess of 50~$\AA^{-1}$. The observations covered 
a spectral range of $\sim$3400 to $\sim$7300~$\AA$. This covers the Rose 
(1984) indices and 25 Lick indices from H$\delta_A$ to the TiO indices 
(for Lick index definitions used see Trager et al. 1998 and Worthey \& 
Ottaviani 1997). The main aim of the P02 study was the measurement of 
Lick indices in metal--rich bulge GCs. Eight such clusters were observed, 
as well as four associated with the halo. In order to ensure a 
representative sampling of the underlying stellar population a number of 
(generally three) individual spectra were obtained, offset from each 
other by a few arcseconds. The bulge spectrum was constructed from 15 different 
bulge fields, many in Baade's Window. Data reduction included calibration to 
the Lick system using 12 Lick standard stars observed on the same observing 
run.

Our second source of Galactic GC data is from Cohen, Blakeslee \& Ryzhov 
(1998; hereafter CBR98). Lick index values were taken from Beasley et al. (2004). 
The data were obtained using the Keck I telescope with a dispersion of 
1.24~$\AA$/pix and achieved a signal--to--noise similar to that of the 
P02 data (i.e. S/N$\sim$50~$\AA^{-1}$). For approximately half the GCs in the sample, 
Lick index measurements were made between 4200~$\AA$ and 6500~$\AA$, 
while in the other half of the sample indices were measured between 
4600~$\AA$ and 6500~$\AA$. The study included twelve GCs from both the halo 
and bulge sub--populations. Table \ref{CaII} shows that the sample contains 
red, blue and intermediate horizontal branch GCs. The combined P02 and CBR98 samples 
include 20 GCs (four of which were observed by both studies). 

Due to the lack of Lick standard star observations, the CBR98 data were 
not calibrated to the Lick system. Fortunately, there are four GCs in 
this data set common to the (fully calibrated) P02 study. The average 
differences between the indices in the common GCs were therefore used 
to calibrate the CBR98 data to the Lick system. The offsets were generally 
found to be $\sim$0.3~$\AA$ in line indices and $\sim$~0.01 
mag in molecular band indices, with no evidence of correlations with index
values. These offsets are consistently smaller than 
the rms scatter in the differences between GCs. However, for Fe5015 an 
offset of 0.46~$\AA$ (with 0.50~$\AA$ rms) was found between the data sets 
(with the CBR98 data exhibiting lower values). We leave interpretation of 
this offset to Section \ref{fitting_proc}. 

As this method of calibrating to the Lick system is far less reliable than 
using Lick standard stars, the index errors for the CBR98 data
were taken to be the rms about the mean offsets for the four common GCs. 
This value is assumed for the error as it consistently exceeds the quoted 
observational error. However, we note that it is the \emph{relative} size of 
index errors that influences the $\chi^2$ fitting procedure outlined in
Section \ref{fitting_proc}.

\subsection{Rose indices and horizontal branch stars}
\begin{figure}
\centerline{\psfig{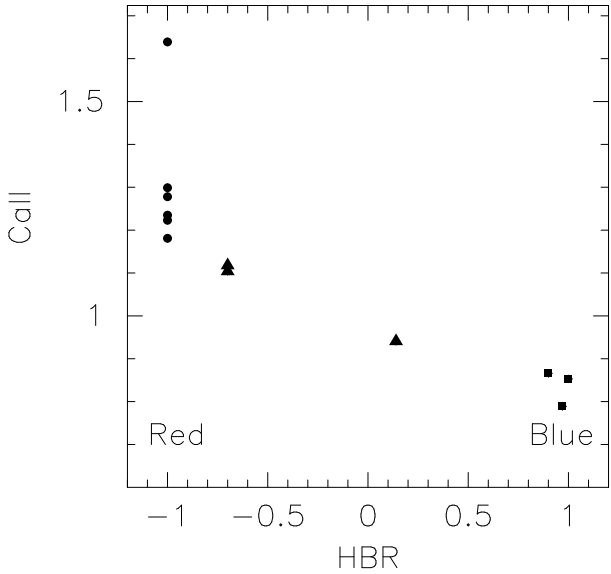}}
\caption{CaII index against horizontal branch ratio (HBR) for the P02
sample. Intermediate horizontal branches (IHBs; triangles) are evident in 
Piotto et al. (2002) colour--magnitude diagrams of these GCs. Pure blue 
and red horizontal branches are shown by squares and circles respectively.
The CaII index provides a good indicator of the HBR.}
\label{CaII_HBR}
\end{figure}

We have measured the CaII index (Rose 1984) from the P02 spectra. 
This index is known to be sensitive to the presence of hot stars 
(Rose 1985) such as those occupying the blue horizontal branch (BHB). 
The measured CaII values are given in Table \ref{CaII} and plotted 
against horizontal branch ratio (HBR) from Harris (1996) in Fig. 
\ref{CaII_HBR}. The three intermediate HBR GCs in this figure are shown 
by Piotto et al. (2002) to possess both red and blue horizontal branch 
stars. We shall therefore refer to such GCs as possessing intermediate 
horizontal branches (IHBs). Fig. \ref{CaII_HBR} suggests that the 
CaII index allows an estimation of the HBR in old stellar populations 
for which CMDs can not be obtained, i.e. from the integrated spectra 
of most extra--galactic systems. Schiavon et al. (2004) detailed an alternative approach to the identification 
of GCs with BHBs, achieved by comparing the ratio of H$\delta_F$ to H$\beta$ to
\emph{their} SSP models. We found the ratio predicted by the Schiavon et al.
(2004) models to be less than the SSP model
predictions used here (see Section \ref{models}) for essentially \emph{all} the GCs in our sample, i.e. the
excess in the ratio in BHB GCs found by Schiavon et al. (2004) is not reproduced 
in the models used here. One possible explanation of this
is that Schiavon et al. use the Jones (1996) stellar library, while the
models used in this work are based on the Lick stellar library (see Section
\ref{models}).

\begin{table}
\footnotesize
\begin{center}
\begin{tabular}{lccr}
\hline
NGC  &Messier &   CaII & HBR\\
\hline
\multicolumn{4}{l}{\bf Puzia et al. (2002) GCs}\\
5927 &  - &  1.223&--1.00\\
6388 &  - &  1.104&--0.70\\
6528 &  - &  1.181&--1.00\\
6624 &  - &  1.235&--1.00\\
6218 & 12 &  0.789& 0.97\\
6441 &  - &  1.118&--0.70\\
6553 &  - &  1.639&--1.00\\
6626 & 28 &  0.866& 0.90\\
6284 &  - &  0.853& 1.00\\
6356 &  - &  1.299&--1.00\\
6637 & 69 &  1.278&--1.00\\
6981 & 72 &  0.941& 0.14\\
\hline
\multicolumn{4}{l}{\bf Cohen et al. (1998) GCs}\\
6205 & 13  &   -   &   0.97\\
6121 & 4   &   -   & --0.06\\
6838 & 71  &   -   & --1.00\\
6341 & 92  &   -   &   0.91\\
6171 & 107 &   -   & --0.73\\
6356 &  -  &   -   & --1.00\\
6440 &	-  &   -   & --1.00\\
6528 &	-  &   -   & --1.00\\
6539 &	-  &   -   & --1.00\\
6553 &	-  &   -   & --1.00\\
6624 &	-  &   -   & --1.00\\
6760 &	-  &   -   & --1.00\\
\hline
\end{tabular}
\end{center}
\caption{CaII index and horizontal branch ratios (HBRs; Harris 1996) for 
Galactic globular clusters.}
\label{CaII}
\normalsize
\end{table}

\subsection{The SSP models}
\label{models}
\begin{figure}
\centerline{\psfig{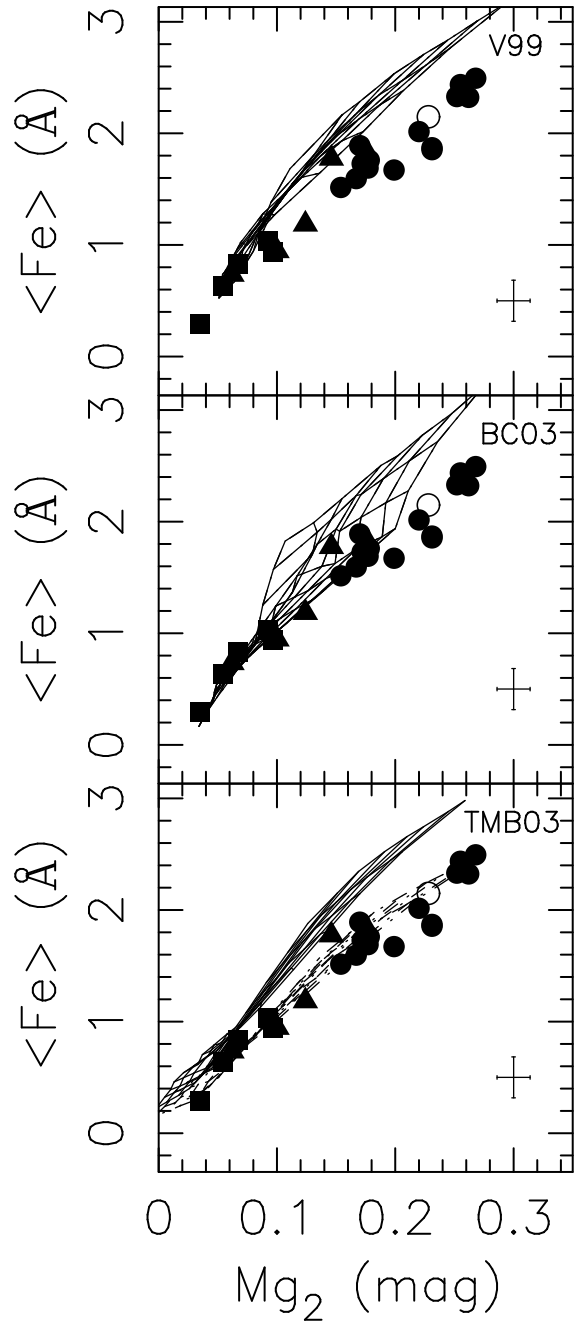}}
\caption{The data of P02 and CBR98 are plotted on SSP 
grids. Symbols represent horizontal branch morphology as in Fig. 
\ref{CaII_HBR}. The average error in each index is indicated in the bottom
right of each plot. Lines of constant metallicity range from [Fe/H]=0.0 down 
to the lowest metallicity modelled (-1.5 for V99 and --2.25 for BC03 and 
TMB03). Lines of constant age are from 1~Gyr up to the oldest population 
modelled (17, 20 and 15~Gyr for V99, BC03 and TMB03 respectively). For 
TMB03 models, solar abundance ratio models are shown as solid lines, while
models with an $\alpha$--element enhancement of +0.3~dex 
are shown as dashed lines. The Galactic bulge 
(open circle) is also shown. The GCs NGC 6356, 6528, 6553 and 6624 appear 
twice in each diagram as both the P02 and  CBR98 values are presented.
The data lie off the V99 and BC03 grids at high index values, as they possess
$\alpha$--element enhancements greater than that modelled.}
\label{grids1}
\end{figure}

The results from three different SSP models are compared in this paper. 
These are:\\

\noindent {\bf Vazdekis (1999; hereafter V99):} The models used here are
those based on the fitting functions of Worthey et al. (1994) and Worthey \&
Ottaviani (1997). These SSP models cover 
the metallicity range --1.68~$\le$~[Fe/H]~$\le$~0.2 and ages from 0.1 
to 17.78~Gyr. They use the isochrones of Bertelli et al. (1994). 
These isochrones include BHB stars in old, low-metallicity populations. 
No account is taken of the variation in abundance ratios in the stars 
used to construct the SSPs. The index values therefore reflect the solar
neighbourhood abundance ratio pattern (see Section \ref{NSAR}), with
$\alpha$--element enhancements falling from $\sim$+0.3~dex at
low metallicities to 0.0~dex at solar metallicity.\\

\noindent {\bf Bruzual \& Charlot (2003; hereafter BC03):} The models used
here are those derived from spectral energy distributions and then
calibrated to the lick system (i.e. from their
lsindex\_sed\_lick\_system files). These SSP models cover 
the metallicity range --2.3~$\le$~[Fe/H]~$\le$~0.4 and ages from 0.1 Myr to
20~Gyr. The models use Padova (1994) isochrones (Alongi et al. 1993; Bressan et al. 1993; 
Fagotto et al. 1994a,b; Girardi et al. 1996). These isochrones include
BHB stars in old, low-metallicity populations. No account is taken of the 
variation in abundance ratios in the stars used to construct the SSPs, so, again, the 
index values reflect the local abundance ratio pattern (see
Section \ref{NSAR}).\\   

\noindent {\bf Thomas, Maraston \& Bender (2003; hereafter TMB03):} These
SSP models cover the metallicity range --2.25~$\le$~[Fe/H]~$\le$~0.67 and 
ages from 1 to 15~Gyr. The models use the isochrones of Cassisi, Castellani 
\& Castellani (1997), Bono et al. (1997) and Salasnich et al. (2000). TMB03 
provides two versions of the Balmer lines in their SSP models; one modelling 
blue horizontal branches in low metallicity populations, the other modelling 
red horizontal branches (RHBs) at all metallicities. TMB03 also adjust index values to take account 
of the variation in abundance ratios in the stars used to construct the SSPs 
using the results of Tripicco \& Bell (1995; see Section \ref{NSAR}). TMB03 
thus provide index values for SSPs with both solar-- and non--solar 
($\alpha$--element enhanced) abundance ratios.\\

We refer the reader to the source papers for further details. Here, we focus on 
obtaining reliable estimates of GC properties from each of the models. We
note that Schiavon et al. (2002a,b) have produced SSP models, based on the
colour-magnitude diagram of 47 Tuc. Unfortunately, these models are
as yet unpublished and are, therefore, not considered in this work.\\

The data of P02 and CBR98 are plotted in the $<$Fe$>$--Mg$_2$ plane in 
Fig. \ref{grids1} where $<$Fe$>$=(Fe5270+Fe5335)/2. Age--metallicity grids 
from the SSP models are also shown. Grid lines for age are shown from 1~Gyr 
up to the oldest age modelled. For metallicity, the grids show the lowest
[Fe/H] modelled up to solar ([Fe/H]=0.0~dex). For TMB03 models, only the
models with $\alpha$--element enhancement of +0.3~dex are shown.
However, it is important to note that the TMB03 models differ from 
those of V99 and BC03 in that TMB03 correct their indices for the
abundance ratio pattern in the stars used to construct the SSP models 
(see Sections \ref{NSAR} and \ref{enhest}). The TMB03 models, therefore, 
reflect a constant $\alpha$--element enhancement at all metallicities, 
while V99 and BC03 SSP models reflect the varying abundance ratio with 
metallicity of the stars used in their construction.

The effect of the varying abundance ratios with metallicity in the V99 and
BC03 models can be seen in Fig. \ref{grids1}. For all GCs, the TMB03 models suggest
a near constant $\alpha$--element enhancement which is consistent with the high resolution
study of Carney (1996). However, in V99 and BC03 plots, while GCs with low 
$<$Fe$>$ and Mg$_2$ (i.e. low metallicity GCs), lie close to, or within, the
grids, those with high index values lie well to the right of the grids. This
is simply the result of low $\alpha$--element enhancements in
high-metallicity, local-neighbourhood stars used to construct the V99 and BC03 
SSP models (e.g. Edvardsson et al. 1993; Gustafsson et al. 1999; Bensby, 
Feltzing \& Lundstr{\o}m 2003).

\begin{figure*}
\centerline{\psfig{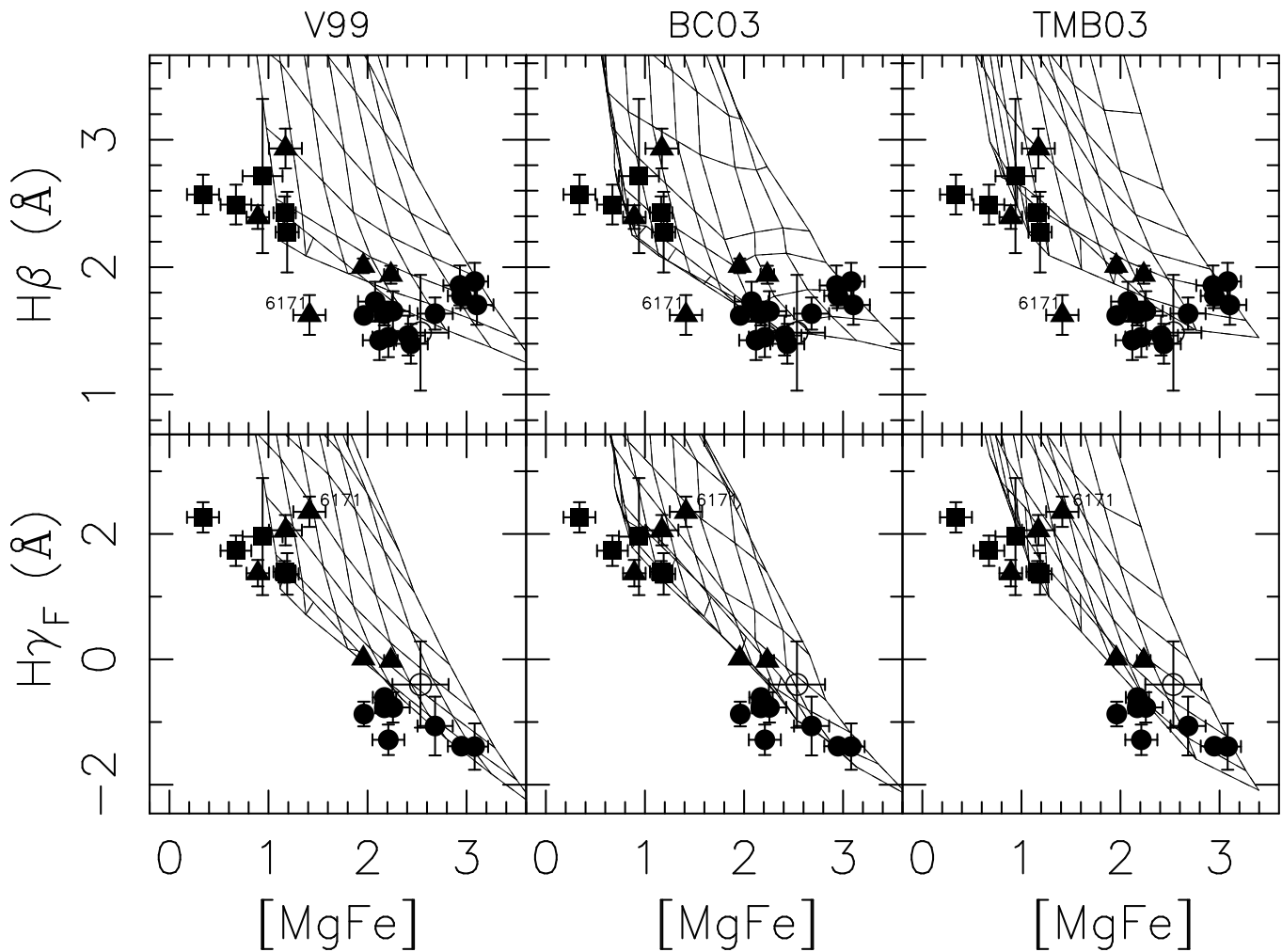}}
\caption{The data of P02 and CBR98 are plotted on H$\beta$--[MgFe] 
and H$\gamma_F$--[MgFe] SSP grids. Symbols are as in Fig. \ref{CaII_HBR}. 
Grids lines in metallicity are from
[Fe/H]=0.0 (solar) down to the lowest modelled (-1.5 for V99 and --2.25 for
BC03 and TMB03) in steps of 0.25~dex. For age, grid lines are 1, 1.5, 2, 3,
5, 8, 12~Gyr as well 
as the highest age modelled (17, 20 and 15~Gyr for V99, BC03 and TMB03 
respectively). There are fewer points in the H$\gamma_F$ plots, as this
index is not measured in half of the CBR98 GCs.
A wide range of ages is suggested by both H$\beta$ and 
H$\gamma_F$ plots, with many GCs falling outside the SSP grids. Large
variations between the suggested ages in H$\beta$ and H$\gamma_F$ plots are
also evident, particularly for NGC~6171 and the bulge (open circle).}
\label{grids2}
\end{figure*}

The variations in abundance ratios severely complicate the process 
of deriving age and metallicity from the SSPs models. In order to overcome 
these difficulties, the [MgFe] index was defined as 
[MgFe]=$\sqrt{Mgb\times(Fe5270+Fe5335)/2}$ (Gonzalez 1993). This combination 
of Lick indices was selected as it is insensitive to enhancement effects. Two 
commonly used Balmer lines (H$\beta$ and H$\gamma_F$) are plotted against 
[MgFe] in Fig. \ref{grids2}. Such two dimensional index plots are commonly 
used to derive ages and metallicities (e.g. Bressan, Chiosi \& Tantalo 1996;
Kuntschner \& Davies 1998;
Longhetti et al. 2000; Kuntschner et al. 2002; Beasley, Hoyle \& Sharples 2002). 
However, inspection of Fig. \ref{grids2} clearly identifies a number of concerns. 
Firstly, in both H$\beta$ and H$\gamma_F$ plots, a significant number of the GCs 
fall below the oldest SSPs, suggesting ages significantly greater than the
age of the universe. A second concern is that a number of GCs suggest young 
($\sim$5~Gyr) ages, in conflict with the known age distribution of these Galactic
GCs (Salaris \& Weiss 2002; Rosenberg et al. 1999). In addition, for many GCs, the 
ages suggested by the H$\beta$--[MgFe] grids differ from those suggested by the
H$\gamma_F$--[MgFe] grids. For example, NGC~6171 lies below the grids in H$\beta$, 
suggesting an age $>$15~Gyr, while in the H$\gamma_F$ plots its position within the 
grid suggests an age $<$5~Gyr. The bulge also suggests a younger age in the 
H$\gamma_F$--[MgFe] plots than in those of H$\beta$--[MgFe]. On the other hand, 
the high metallicity GCs, NGC~6528 and 6553 (with [MgFe]$\simeq$3.0~\AA),
suggest \emph{older} ages in the H$\gamma_F$--[MgFe] plots than in the  H$\beta$--[MgFe] plots.
Therefore, the ages derived from Fig. \ref{grids2} are sometimes highly dependent on the choice
of Balmer line, and are inconsistent with the known age distribution of these 
Galactic GCs, regardless of which Balmer line is used. Consequently, we 
conclude that 2-dimensional index plots, such as those in Fig. \ref{grids2}, 
are unreliable for the measurement of ages in globular clusters. In the following, 
we outline a method that is robust with respect to such problems.

\section{Fitting data to SSP models}
\label{fitting}
The method of measuring the derived parameters for GCs used in this work
is a $\chi^2$--minimisation technique which involves fitting 
as many Lick indices as possible. The justifications for this are:
\begin{itemize}

\item By fitting as many indices as possible the procedure makes maximum 
use of the available data. It should be noted that,
while individual indices are degenerate with respect to the derived
parameters (age, metallicity and $\alpha$--element abundance ratios), most 
indices contain \emph{some} information regarding each of these parameters.

\item The derived parameters are less prone to uncertainties in data 
reduction, e.g. flux calibration error, stray cosmic rays, skyline 
residuals, velocity dispersion error, errors in conversion to the Lick 
system, etc.

\item The procedure is also particularly useful for spectra affected 
by specific modelling problems, such as horizontal branch morphologies 
not matching those modelled in the SSPs. This is because the affected 
indices (mainly H$\beta$ in the case of horizontal branch morphology 
problems) can be identified and, if necessary, omitted from the fitting 
procedure with only a modest increase in overall error (see Proctor \& 
Sansom 2002; hereafter PS02). Indeed, using this technique, 
the derived parameters are less prone to modelling uncertainties in general.
\end{itemize}

\subsection{Modelling the effects of non--solar abundance ratios}
\label{NSAR}
In this work we use SSP models to derive estimates of age, [Fe/H] 
and the abundance ratio of a number 
of key, mainly $\alpha$--, elements (parameterised by [E/Fe] -- see PS02). 
TMB03 models include SSPs of solar abundance ratio (i.e. 
[E/Fe]~=~0.0) as well as [E/Fe]~=~--0.3, +0.3, +0.5. These models were
constructed using the results of Tripicco \& Bell (1995; hereafter TB95)
by the method first outlined in Trager et al. (2000; hereafter T00).
V99 and BC03 SSP models, 
on the other hand, have abundance ratios that vary with metallicity, as their
models reflect the abundance ratios of the stellar calibrators used to construct 
the models. We have calculated indices for V99 and BC03 SSP models for
abundance ratios [E/Fe]=--0.3,+0.3 and +0.6, using the results of TB95. Two
methods were considered. The first, (the T00 method), 
is best suited to low metallicity ([Fe/H]$<$0.0) populations. The second, 
described in PS02 (the `Fe-- method'), is better suited to high metallicity 
populations ([Fe/H]$\gtrsim$0.0).
Both of these methods are based on the assumption of a fixed fractional
contribution to the total luminosity of each the three stellar types modelled 
by TB95. 

The fractional change in index values for a change of +0.3~dex in [E/Fe] are
listed in Table \ref{fracts} for both methods. TB95 did not model H$\delta$
and H$\gamma$ indices, so these indices are assumed to be unaffected
by changes in abundance ratios\footnote{We note that,
Thomas, Maraston \& Korn (2004) have recently estimated the sensitivity of the
H$\delta$ and H$\gamma$ indices to abundance ratios.  However, they find
that; {\it ``{\ldots} the sensitivity of H$\gamma$ and H$\delta$ to
$\alpha$/Fe disappears at low metallicity and is relatively small around
solar metallicity {\ldots} ''}.}.

It is important to note that the two methods assume different
chemical compositions, with the T00 method assuming a fixed [Z/H] (overall
metallicity) as [E/Fe] changes, while the PS02 method assumes fixed [Fe/H].
Unless otherwise stated, the derived parameters from these SSP models quoted 
in this work were derived using the T00 method. Linear interpolation (in 
grid steps of 0.025~dex) allowed us to estimate intermediate values of the
derived parameters.

\begin{table}
\footnotesize
\begin{center}
\begin{tabular}{lrr}
\hline
Index  &PS02 & T00 \\
       &method&method\\
\hline
H$\delta_A$&        --    &      --  \\
H$\delta_F$&        --    &      --  \\
CN1       &       0.058   &    0.029 \\
CN2       &       0.071   &    0.032 \\
Ca4227    &      -0.092   &   -0.272 \\
G4300     &       0.097   &    0.053 \\
H$\gamma_A$&        --    &      --  \\
H$\gamma_F$&        --    &      --  \\
Fe4383    &      -0.029   &   -0.132 \\
Ca4455    &       0.185   &    0.012 \\
Fe4531    &       0.054   &   -0.077 \\
Fe4668    &       0.368   &    0.032 \\
Hbeta     &       0.031   &    0.031 \\
Fe5015    &       0.059   &   -0.077 \\
Mg1       &       0.048   &    0.024 \\
Mg2       &       0.048   &    0.018 \\
Mg b      &       0.315   &    0.207 \\
Fe5270    &      -0.012   &   -0.127 \\
Fe5335    &      -0.083   &   -0.196 \\
Fe5406    &      -0.058   &   -0.172 \\
Fe5709    &       0.015   &   -0.068 \\
Fe5782    &       0.049   &   -0.159 \\
Na D      &       0.258   &    0.025 \\
TiO1      &       0.002   &    0.001 \\
TiO2      &       0.002   &   -0.001 \\
\hline
\end{tabular}
\end{center}
\caption{Fractional change in indices for an [E/Fe] change of +0.3~dex,
calculated from TB95 (see PS02). TB95 did not model H$\delta$ and H$\gamma$ indices. 
These indices are
therefore assumed to have no sensitivity to [E/Fe]. It should be noted that
application of the two methods differs as they describe 
different chemical compositions (see PS02).}
\label{fracts}
\normalsize
\end{table}

\subsection{Fitting procedure}
\label{fitting_proc}
As our aim is to fit as many indices as possible, we began by obtaining
the best fit (by $\chi^2$--minimisation) to all 25 observed Lick indices 
in log(age), [Fe/H] and [E/Fe] parameter space. For the TMB03 SSP models, 
fits were obtained to both the RHB and BHB models. 

\begin{figure*}
\centerline{\psfig{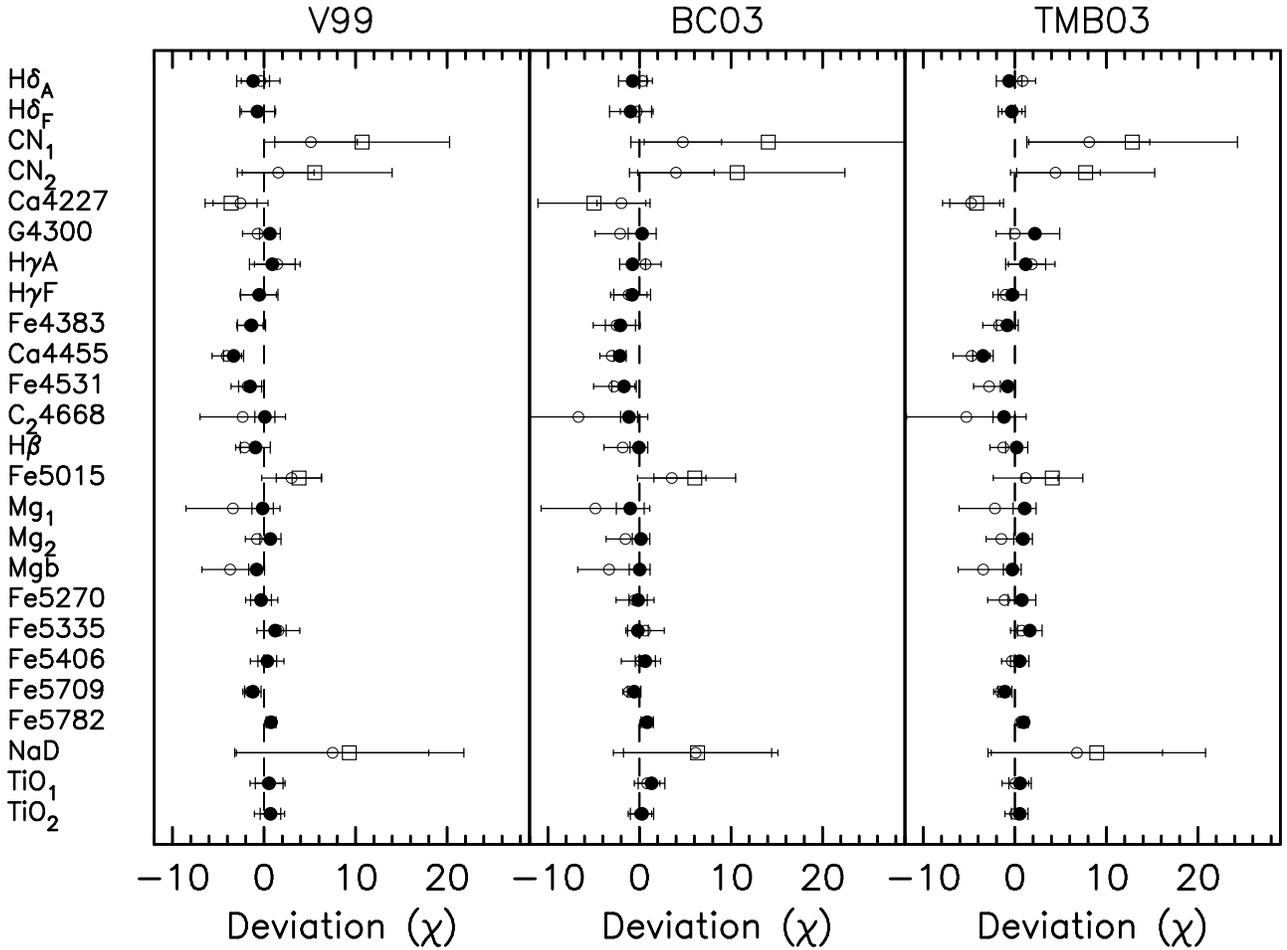}}
\caption{The average deviation in units of error (i.e. $\chi$) for
GCs in the P02 sample. Indices are listed in wavelength order.
Error bars represent the rms scatter in the deviations. 
Open circles show the results of including all 25 indices in the fitting
procedure. Solid circles represent the deviations when poorly fitting indices (open
squares) are omitted from the fitting procedure (see text).}
\label{chiplot}
\end{figure*}

Residuals to the best fit (observed value minus best fit value expressed 
in terms of index errors, i.e. $\chi$) are summarised in Fig. \ref{chiplot}. 
This figure shows the average and rms deviations from the best fit values for 
all indices of the P02 sample, and for all three sets of SSP models. It is 
clear from 
this figure that a number of indices are poorly fit by all three sets of 
SSP models. The poorest fits were obtained for the CN indices and NaD. The 
NaD index is known to be severely affected by interstellar absorption. This 
index was therefore excluded from the 
fitting procedure. On the other hand, the CN indices, which were also 
found to be enhanced by Trager (2004), are extremely sensitive to nitrogen 
abundance (TB95) which is known to be higher in Galactic GCs than in field
and halo stars (e.g. Li \& 
Burstein 2003). The large positive residuals present in the fits to GCs 
(but not evident in the fit to the P02 bulge data) is therefore consistent 
with an [N/Fe]~$>$~0 in GCs. Indeed, TMB03 showed that SSP models in which N
is independently enhanced with respect to other light elements can provide
excellent fits to these indices. The fitting procedure has therefore successfully 
identified this abundance anomaly. Rather than introduce N abundance as another free
parameter in our modelling, the CN indices were excluded from 
the fitting procedure in the following. 

As the CN and NaD indices were removed form the fitting procedure, the fits of 
other indices (particularly the enhancement sensitive indices -- C$_2$4668,
Mg$_1$, Mg$_2$ and Mgb) are generally
improved (see Fig. \ref{chiplot}). However for two indices (Fe5015 and Ca4227) this 
was not the case. For Fe5015 we recall that this index showed a large variation 
between the two samples for the four GCs in common to P02
and CBR98. The 
positive deviation in Fig. \ref{chiplot} is consistent with the assumption that 
P02 overestimate the Fe5015 index values. In this case the fitting procedure therefore 
appears to have identified a problem in calibration to the Lick system. Consequently, Fe5015 
was excluded from the fitting procedure.

The Ca4227 index was also shown by TB95 to be sensitive to nitrogen abundance (with 
increased N abundance resulting in a reduction in the Ca4227 line--strength). Its 
negative deviation in Fig. \ref{chiplot} is therefore consistent with the 
nitrogen overabundance in GCs. However, Ca4455 also shows a  negative deviation. 
This weakness in the calcium lines has previously been observed in GCs (e.g.
Maraston et al. 2003) 
and galaxies (e.g. Worthey 1992; Vazdekis et al. 1997; Trager et al. 1998;
PS02), as well as in the near-infrared CaT index (e.g. Saglia et al. 2002).
It is therefore unlikely to be a result of problems in
calibration to the Lick system. In addition, since TB95 showed Ca4455 to be 
insensitive to calcium abundance, it is unlikely that this is the effect 
of a variation in [Ca/Fe]. Finally, we note that the indices
either side of Ca4455, particularly Fe4383 and Fe4531, 
also possess consistently negative deviations, suggesting that there is some
systematic discrepancy between models and observations at these wavelengths.
We therefore conclude that the poor fit of Ca4455 is most likely the result of 
problems in the modelling of this index in SSP models. 

Despite its relatively large
deviations in Fig. \ref{chiplot}, exclusion of Ca4455 had negligible
impact on the results of the fitting procedure, with 85\% of the fits 
finding \emph{exactly} the same best fit, while the remaining 15\% differed 
by at most 2 grid steps (amounting to $\lesssim$0.05~dex in any derived 
parameter). To be consistent with our aim of fitting 
as many indices as possible we therefore decided to include this index 
in the fitting procedure.

Having considered the average and rms deviations to the best fits, we turn our
attention to the deviations of indices in individual GCs. For the most part,
no single index dominated the deviations in any given GC. The main exception to
this being H$\beta$ in seven GCs fitted using either V99 or BC03 SSP models, as
well as the TMB03 RHB models. For the three of these aberrant GCs in the P02 data,
CaII values are all below 1.15, indicating that these are all either BHB or
IHB GCs. The decision was therefore taken to exclude
H$\beta$ from all GCs in the P02 sample with CaII below this value, i.e. in all
the BHB and IHB GCs. For the four
remaining CBR98 GCs with deviations dominated by H$\beta$ deviations, the index was
also omitted from the fitting procedure. It is worth noting that these four GCs are 
also those in the CBR98 data with HBR values $>$0.9 and therefore constitute all the 
BHB and IHB GCs in the CBR98 sample.

The decision to exclude certain H$\beta$ values is supported by the finding of PS02 that 
the exclusion of Balmer lines from the fitting procedure has relatively little impact 
on the derived parameters, as long as a sufficiently large number of indices are
included in the fitting procedure. The results from TMB03 SSP models are taken to be 
those from the RHB fits in GCs with HBR=--1.0, and for 
BHB fits for GCs with HBR$>$--1.0 (i.e. IHB and BHB GCs). 

Finally, the deviations from the best fits of TiO indices in 
NGC~6626 and G4300 in NGC~6637 and NGC~6981 were also found 
to be large.  Consequently, these indices were excluded from the fitting of the 
respective GCs. The final deviations from the best fits to the P02 data are 
shown in Fig. \ref{chiplot}.

A similar process was carried out for the CBR98 data. However,
these data spanned a narrower wavelength range and, consequently, it was only
possible to include between 8 and 16 indices in the fits (compared to
between 18 and 20 using P02 data). For common indices the results 
were similar to those for the P02 data, with the exception of the TiO 
indices which were found to give poor fits in the CBR98 data. TiO$_1$
and TiO$_2$ were therefore excluded from the fitting of the CBR98 GCs.
H$\gamma$ indices in NGC~6171 were also found to possess large residuals to
the best fits (see Fig. \ref{grids2}). These indices were therefore omitted
from the fitting of this GC.

The best fits of the P02 data to all three SSP model sets exhibited average 
reduced--$\chi^2$ values of $\sim$3.5. This suggests that either
reduction errors exist that have not been accounted for, and/or that significant errors 
are present in the SSP models. 
Indeed, the typical reduced--$\chi^2$ values for our fits to CBR98 data (for
which the rms scatter in the comparison of 4 GCs common to both P02 and
CBR98 was used as the error)  were $\sim$1.5, suggesting that both effects
noted above are present. However, whatever the case, our use
of a large number of indices in the fitting procedure ameliorates the
effects of such problems on the estimates of derived parameters. 
It is also worth noting that, since the reduced $\chi^2$ of the best fits 
were significantly greater than 1.0, results for lower signal-to-noise spectra 
can be expected to be similar. However, from previous experience (e.g.
Proctor et al. 2004), we find that a minimum signal-to-noise of 20 \AA$^{-1}$ 
is required for the technique to obtain robust results.

In order to test the robustness of the fitting procedure, fits were obtained
for each of the GCs with only eight indices included in the fitting
procedure (we recall that between eight and 20 indices were included in the
final fits above, with the majority of GCs fit by 16 or more indices).  
The eight indices included in these fits were H$\beta$, Mg$_1$, Mg$_2$, Mgb,
Fe5270, Fe5335 Fe5406 and Fe5709. 
The rms difference between the derived parameters from these fits and those from 
the final fits above were 0.103~dex, 0.144~dex and 0.087~dex in log(age),
[Fe/H] and [E/Fe] respectively.

In summary, the $\chi^2$--minimisation technique has allowed
us to identify problems with NaD (probably caused by interstellar
absorption), the known nitrogen abundance
anomaly affecting CN$_1$, CN$_2$ and (possibly) Ca4227, problems in
calibration to the Lick system (Fe5015 in the P02 data and TiO indices in
CBR98), as well as individual indices
suffering unexplained reduction errors. This technique therefore holds a
great deal of promise in the investigation of integrated stellar populations such as
extra-galactic globular cluster systems.

\section{Comparing derived parameters to literature values}
In this section we review the derived parameters (age, [Fe/H], [Z/H] 
and [E/Fe]) obtained from the fitting procedures described in the previous 
section. These are denoted with the subscript SSP.
The results are compared to those obtained by other techniques (such as
fitting to the H$\beta$--[MgFe] and H$\gamma_F$--[MgFe] grids shown in Fig.
\ref{grids2}) and to values from the literature. We wish to emphasise that the 
values of
the derived parameters were not considered in the fitting procedure outlined
in the previous section, i.e. the `best' fits were arrived at without 
considering the derived parameters that they generate. 

Errors in the derived parameters were estimated using Monte--Carlo type 
realisations of the best fit SSP models perturbed by the index error estimates. 
However, this method produces errors in the derived parameters based purely 
on the estimates of observational error, they do
not include any allowance for modelling errors. We estimate more realistic 
errors in Section \ref{discussion}.

Six GCs (NGC~6121, 6205, 6218, 6341, 6626 
and 6981) were found to possess [Fe/H]$_{SSP}$ below the minimum modelled by 
V99. These GCs have therefore been omitted from presentations of V99 results.
NGC~6341 also fell  below the minimum [Fe/H]$_{SSP}$ in BC03 and
TMB03 SSP models. This GC has therefore been omitted from presentations of
BC03 and TMB03 results.

\subsection{Age estimates}
\label{age}
Before discussing the results of the age determinations as a whole, it 
is first necessary to consider the method employed in the fitting of the 
high metallicity GCs NGC~6528 and NGC~6553, as these possess metallicities 
close to solar, i.e. in the range affected by the changing sensitivity of 
isochrones to non--solar abundance ratios (see PS02). 
It was shown in PS02 and Thomas \& Maraston (2003) that the
main effect of this on the values of derived parameters is in age. We
confirm this finding using the GC data of both P02 and CBR98, as values 
of [Fe/H]$_{SSP}$ and [E/Fe]$_{SSP}$ are consistent between the T00 and PS02 
methods\footnote{We note that the application of the PS02 method to TMB03 solar
abundance ratio SSP models causes a small (unavoidable) internal inconsistency 
in these models, as the TMB03 models are already corrected for the local 
abundance ratio pattern \emph{using a method similar to T00}. However, the 
effects of this are expected to be negligible, as the local abundance ratio,
for which TMB03 compensate in generating their solar abundance ratio SSPs, is 
close to the solar value for SSPs with [Z/H]$\sim$0.0~dex.}
(within errors) for both GCs in both 
studies. Values of age, on the other hand, show significant differences. 
The average age and rms value (from both GCs compared to three sets of SSP
models) using the PS02 method is 10.3$\pm$1.5~Gyr, 
while values derived by the T00 method were 7.8$\pm$3.3~Gyr with one 
value as low as 4~Gyr. For the bulge, the PS02 method finds ages of
6.0$\pm$0.8~Gyr while the T00 method gives 3.8$\pm$0.4~Gyr.
The PS02 method therefore produces age estimates that are higher than those
derived by the T00 method (as noted in PS02 and Thomas \& Maraston 2003), and 
more consistent with the known old-ages of GCs  (Salaris \& Weiss 2002;
Rosenberg et al. 1999) and the bulge (e.g. Rich 1999). We have therefore elected 
to use the derived values obtained by the PS02 method for NGC~6528, 6553 and 
the bulge in the following analysis.

\begin{figure*}
\centerline{\psfig{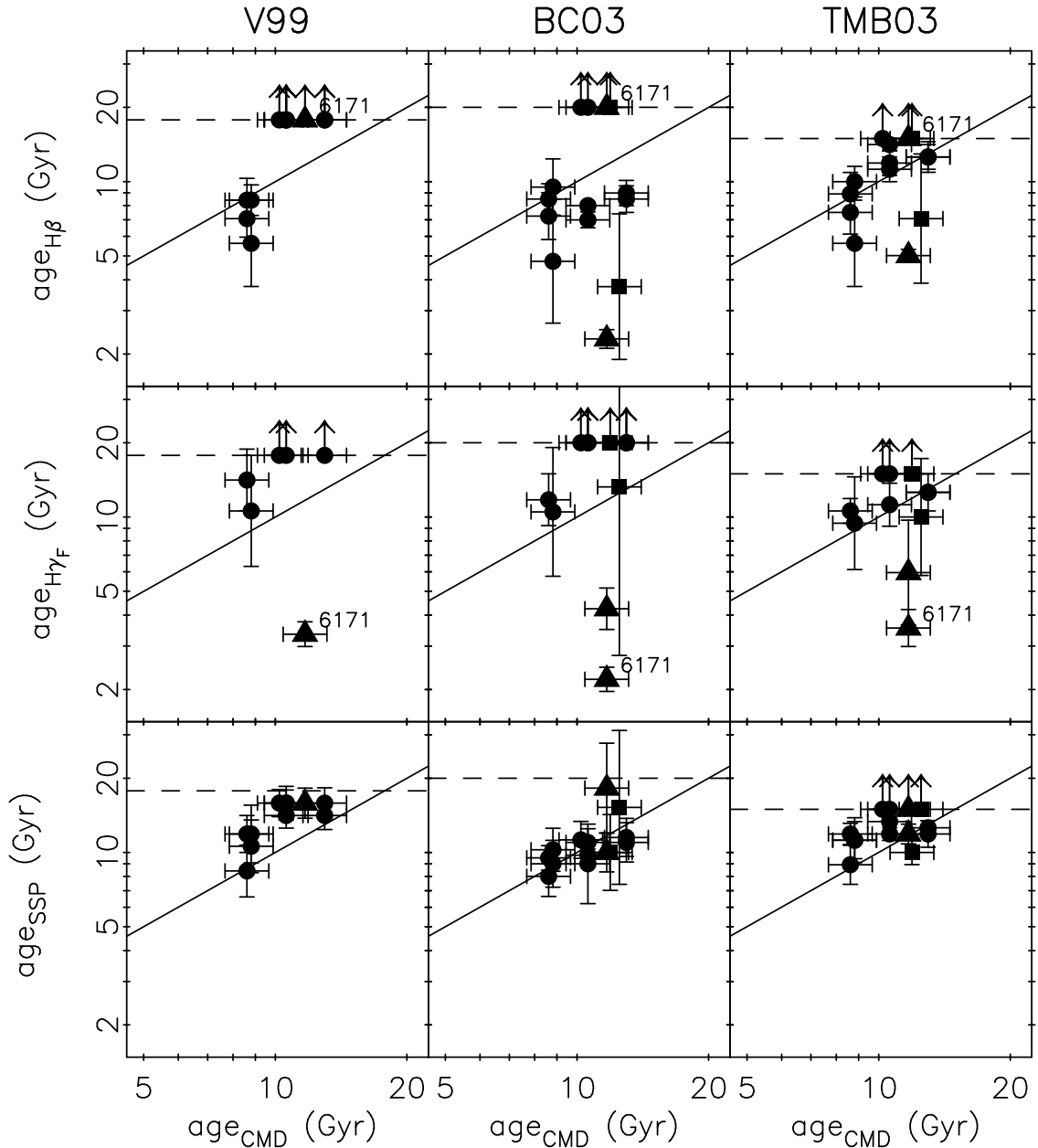}}
\caption{Ages derived from H$\beta$--[MgFe], H$\gamma_F$--[MgFe]
and the $\chi^2$--minimisation technique (top, middle and bottom plots
respectively) are compared to values
derived from CMDs. Values of age$_{CMD}$ are
from Salaris \& Weiss (2002) and Bruzual et al. (1997). One-to-one lines are
shown in each plot (solid lines), as well as the highest SSP age modelled (dashed
line). Points that lie off the Balmer-[MgFe] grids are plotted at the
highest age modelled and identified with an arrow. GCs for which the
$\chi^2$--minimisation technique found ages equal to the oldest modelled are
also identified by an arrow. While very poor agreement is found with CMD ages for
the  Balmer-[MgFe] grid methods, relatively good agreement is found using the
$\chi^2$--minimisation technique. Note also the age variations between the
Balmer--[MgFe] grid methods.}
\label{age_meth}
\end{figure*}

For the four GCs common to the P02 and CBR98 studies (NGC~6356, 6528, 6553 
and 6624) we find an average age offset of only 0.02~Gyr and an 
rms scatter of  1.7~Gyr about a one--to--one relation, although we note
that the two studies are not fully independent, as the CBR98
data were calibrated to the Lick system using a comparison to the P02 data
(Section \ref{data}). 

Ages measured from the SSP models for 10 GCs are compared with published 
ages derived from colour-magnitude diagrams (CMDs) in Fig. \ref{age_meth}. 
CMD-based values are taken from Salaris \& Weiss (2002) for NGC~6121, 6171, 6205, 
6218, 6356, 6624, 6637 and 6838. For NGC~6528 and 6553 we use the Bruzual et
al. (1997) result that these GCs are $\sim$2~Gyr younger than 47 Tuc, which 
is given as 10.7~Gyr old in Salaris \& Weiss. We therefore assume an age
of 8.7~Gyr for NGC~6528 and NGC~6553.

Fig. \ref{age_meth} shows GC ages derived in three ways. The top plots show
ages derived by interpolation of the H$\beta$--[MgFe] grids (Fig. \ref{grids2}).
The tendency for GCs to fall below the grids in Fig. \ref{grids2} is 
reflected by arrows indicating ages
greater than or equal to that modelled. The extremely young ages (3--5~Gyr) suggested 
for some GCs (primarily IHB and BHB GCs) is also evident. Indeed, there is
generally poor agreement with ages derived from CMDs.
The middle plots show ages estimated by interpolation of the  H$\gamma_F$--[MgFe] 
grids (Fig. \ref{grids2}). Again, a large scatter in ages, and poor agreement 
with ages derived from CMDs are found. The inconsistency of age estimates
derived from comparison to the H$\beta$--[MgFe] and H$\gamma_F$--[MgFe]
grids, noted in Section \ref{models}, is evident, particularly for NGC6171,
as well as NGC~6528 and 6553 (both with age$_{CMD}<$10~Gyr). This emphasises the need for
a method of age measurement that is robust to both observational errors
\emph{and} modelling deficiencies.

In the bottom plots the results of our $\chi^2$--minimisation technique are
shown. In this case, no GCs are found to possess ages less than 8~Gyr,
including IHB and BHB GCs which exhibit old ages. In addition, and despite the
narrow age range suggested by CMD values, we find correlations between
log(age)$_{SSP}$ and log(age)$_{CMD}$ at 95\%, 95\% and 90\% confidence
levels for V99, BC03 and TMB03 SSP models respectively (the correlation in TMB03
results are adversely affected by the relatively low upper age limit of the
models). The $\chi^2$--minimisation technique therefore produces age estimates 
that are in reasonable agreement with values derived from CMDs in both 
absolute and relative terms. We therefore conclude that this
technique is superior to the method of deriving ages from
2-dimensional index plots such as Fig. \ref{grids2} and is,
consequently, well suited to the estimation of ages in extra-galactic GC systems.

\begin{figure*}
\centerline{\psfig{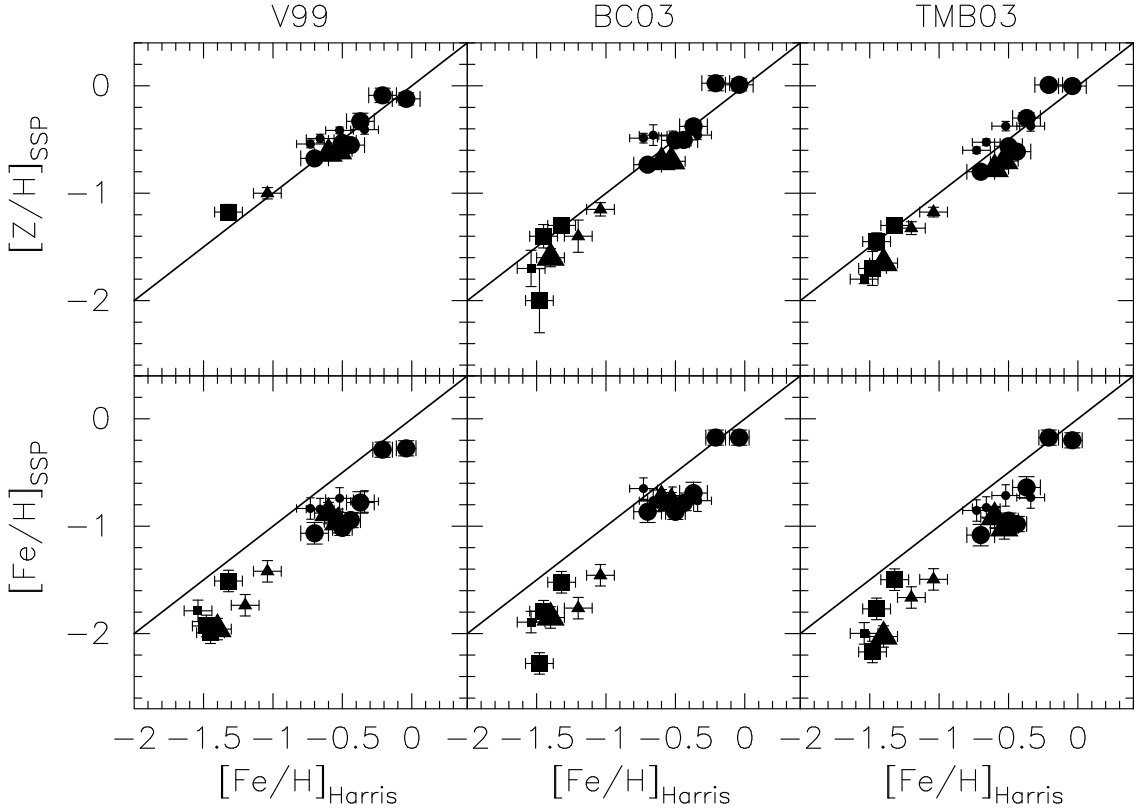}}
\caption{Metallicities ([Fe/H] and [Z/H]) derived from SSP models are compared 
to the [Fe/H] values given in Harris (1996). Symbols shapes represent horizontal 
branch morphology as in Fig. \ref{CaII_HBR}. Symbol size represents GCs from P02 
(large symbols) and CBR98 (small symbols). A good agreement is achieved between
the SSP derived value of [Z/H]$_{SSP}$ and [Fe/H] from Harris (1996) for all
three sets of SSP models.}
\label{harris}
\end{figure*}

\subsection{Metallicity estimates}
\label{met}
In this section we compare the metallicities derived from SSP models 
([Fe/H]$_{SSP}$ and [Z/H]$_{SSP}$) with values published in the literature.

For the four GCs common to the P02 and CBR98 studies (NGC~6356, 6528, 6553
and 6624) we find an average [Fe/H] difference of only 0.01~dex and an
rms scatter of 0.08~dex about a one--to--one relation, although we again note
that the two studies are not fully independent, as we have calibrated the
CBR98 data to the Lick system using a comparison to the P02 data. Given the
good agreement between studies average values are presented in the following.

In their analysis of P02 data, TMB03 found a good correlation between [Z/H] 
and the [Fe/H] values of Harris (1996). A comparison of our [Fe/H]$_{SSP}$ and 
[Z/H]$_{SSP}$ measures, from V99, BC03 and TMB03 SSP models, to the 
values of [Fe/H] of Harris (1996) are shown in Fig. \ref{harris}. The 
same good agreement between [Z/H]$_{SSP}$ and the [Fe/H] values of Harris 
(1996) is found for all three sets of SSP models, with rms variations about
a one-to-one lines $\lesssim$0.2~dex. However, we note that
Beasley et al. (2002) find good agreement between [Fe/H] derived from SSP
models and values from CMDs for GCs in the Large Magellanic Cloud. 
We therefore simply conclude that the methods used in this work (and TMB03) are
well suited to the study of metallicities in extra-galactic GC systems, 
as they produce excellent measures of \emph{relative} metallicities. They also 
successfully recover the actual [Fe/H] value assigned to the isochrone 
that best fits the observed population -- \emph{if} we assume that 
[Z/H]$_{SSP}$ reflects the [Fe/H] of the isochrone. However, this
result suggests that there are still some problems in the calibration of
either the P02 indices (which we used to calibrate the 
CBR98 indices to the Lick system) or the metallicity scale in SSP models.

\subsection{Abundance ratio estimates}
\label{enhest}
In this section we compare the values of [E/Fe]$_{SSP}$, derived by the
$\chi^2$--minimisation technique, with values of
enhancements in $\alpha$--elements published in the literature.

It is important to bear in mind that, for V99 and BC03 SSP models, the values 
of [E/Fe] output from the fitting procedure are enhancements \emph{relative 
to that of 
the best fit SSP models} (see PS02, TMB03 and Proctor et al. 2004 for more
complete discussions). Consequently, since the SSP models are constructed
using stars in the solar neighbourhood, account must be taken of the pattern of
enhancements in local stars (see Section \ref{NSAR}). 
TMB03 have adjusted their indices for the [E/Fe] inherent to local stars 
such that [E/Fe] is quoted with respect to solar at all metallicities. The 
TMB03 SSP models therefore require no further correction. However, for V99 
and BC03 the output values of [E/Fe]$_{SSP}$ need to be corrected for this 
effect. In this work we assume a pattern
in local abundance ratios (and hence in the SSP models) similar to that 
assumed by TMB03. The [E/Fe] of the SSP models is assumed to rise from 
0.0~dex at [Fe/H]$_{SSP}$=0.0~dex to [E/Fe]=+0.25~dex at
[Fe/H]$_{SSP}$=--1.0~dex and to remain at this values for all
[Fe/H]$_{SSP}<$--1.0~dex (e.g. Edvardsson et al. 1993; Gustafsson et al.
1999; Bensby, Feltzing \& Lundstr{\o}m 2003).

\begin{figure}
\centerline{\psfig{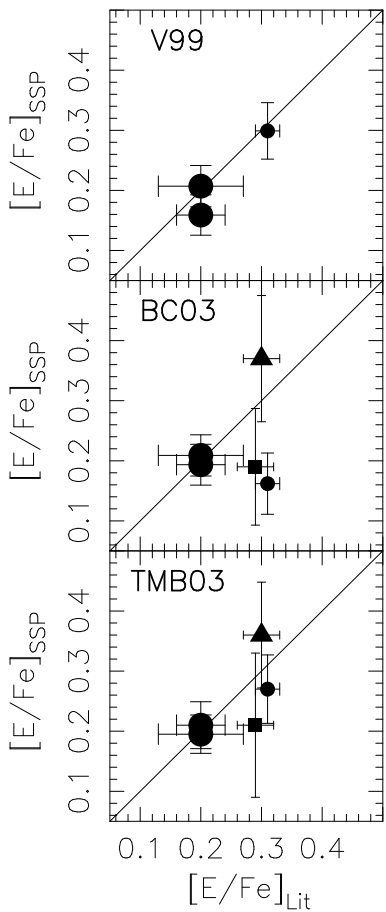}}
\caption{A comparison of [E/Fe] derived from SSP models to those
from the literature. Symbols shapes represent HBR as in Fig. \ref{CaII_HBR}. 
Symbol size represents GCs from P02 (large symbols) and CBR98 (small symbols). 
The one-to-one lines are also shown. The agreement between the two
measurements is good.}
\label{pub_enh}
\end{figure}

A comparison of $\alpha$--enhancements from the literature to [E/Fe]$_{SSP}$
for five GCs (NGC~6121, 6205 and 6838 from Carney 1996, NGC~6528 from
Carretta et al. 2001 and NGC~ 6553 from Cohen et al. 1999) is shown in
Fig. \ref{pub_enh}. For the sake  of consistency with Carney (1996), we use the 
average of the Si, Ca and Ti abundances to calculate $\alpha$--element enhancements 
from the latter two studies. The
agreement between [E/Fe]$_{SSP}$ and the published value of [Si,Ca,Ti/Fe]
for these five GCs is good. We therefore conclude that the $\chi^2$--minimisation 
technique provides an excellent technique for estimating abundance ratios in
extra-galactic GC systems.

\section{Discussion}
\label{discussion}
\begin{figure*}
\centerline{\psfig{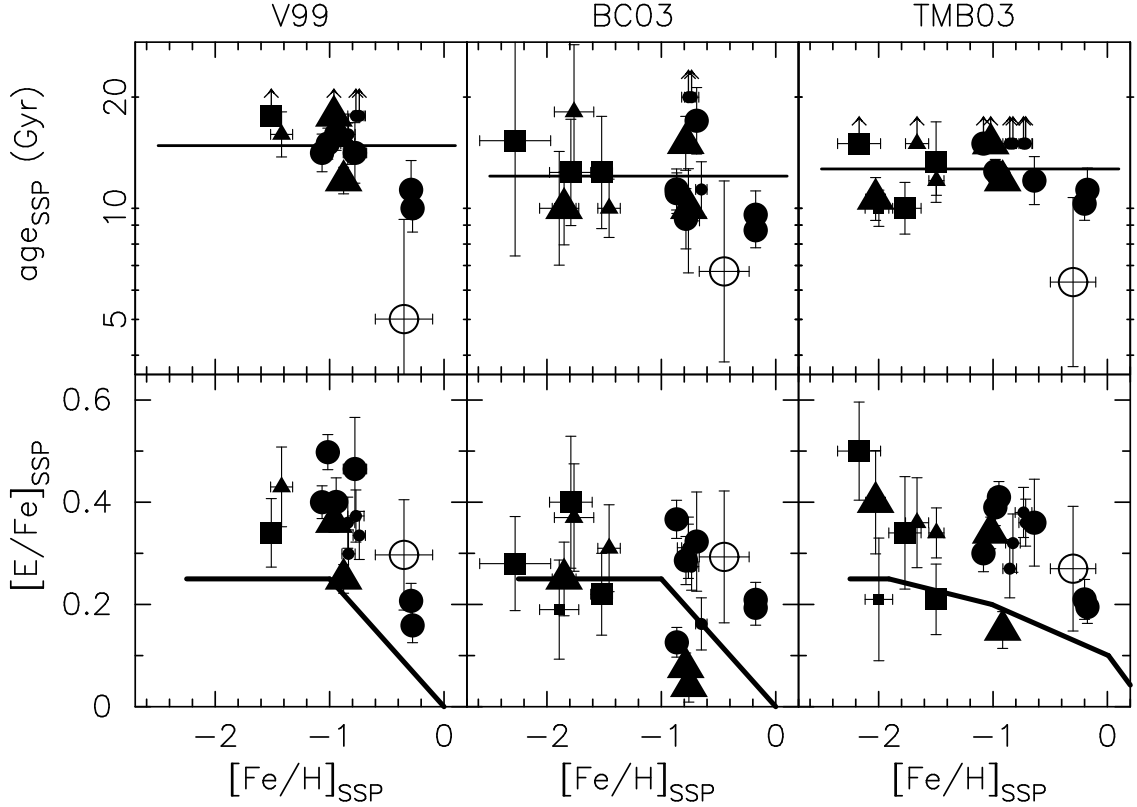}}
\caption{The values of age and local abundance ratio corrected
[E/Fe] are plotted against [Fe/H] derived from SSP models. Symbol shapes
represent horizontal branch morphology as in Fig. \ref{CaII_HBR}. Symbol 
size represents GCs from P02 (large symbols) and CBR98 (small symbols).
Values for the bulge are also shown (open circles). There are no 
discernable differences between the results for the two datasets. Plots 
of age against [Fe/H] (upper) include the average age of the sample 
as horizontal lines. The $\chi^2$--minimisation technique finds GCs to be 
consistently old. The [E/Fe] plots (lower) show the assumed local abundance 
ratio pattern as solid lines. GCs appear slightly enhanced in $\alpha$--elements 
compared to the SSP models at all metallicities. The bulge exhibits a similar 
[E/Fe] to the GCs.}
\label{agez}
\end{figure*}

Given that our aim is to define a procedure for measuring the
derived parameters of extra-galactic GCs, it is clearly useful to
ask: What conclusions would we have drawn from the data as a whole had it 
been an extra-galactic GC system? Can this technique help us in the 
investigation of such systems? To this end, the results of our determinations 
of the derived parameters for the 19 GCs studied are plotted in Fig. \ref{agez}. 

First, we note that the fitting procedure outlined in Section
\ref{fitting_proc}
identified the known abundance anomaly affecting CN indices (assumed to be
the result of nitrogen abundance enhancement). The capability of the
technique to highlight such abundance variations could clearly provide
useful insights into extra-galactic GC systems.

With regard to age estimates, it was shown that the $\chi^2$--minimisation
technique provides values that are in good agreement with CMD derived ages 
in both an absolute and relative sense. The average values of age ($\pm$ 
the rms scatter) were found to be 13.1$\pm$2.3, 12.2$\pm$3.3 and 
12.7$\pm$1.9~Gyr for V99, BC03 and TMB03 respectively. 
Therefore, all three SSP models used indicate an \emph{old} population of GCs. 
Given that at least some of the scatter about these averages is the result
of real variations in GC age, the rms scatters about the average values of 
age indicate that we can safely assume an upper limit of 0.1~dex on
errors in individual log(age) estimates. Such a value is also consistent
with the rms deviations found in the test of robustness detailed in Section
\ref{fitting_proc}.

The metallicity determinations for all three SSP models show a broad range 
of sub-solar metallicities, which are in good agreement with the literature 
values (see Section \ref{met}). The scatter about the one-to-one
relations in the top plots of Fig. \ref{harris} indicate that we can safely 
assume an upper limit of $\pm$0.1~dex on errors in individual [Fe/H] and [Z/H] 
estimates. This value is again consistent
with the rms deviations found in the test of robustness detailed in Section
\ref{fitting_proc}.

The [E/Fe]$_{SSP}$ estimates for the five GCs with literature values for 
$\alpha$-enhancements are in good agreement with the literature values (Section
\ref{enhest}), again in both an absolute and relative sense. The average 
and rms values of [E/Fe]$_{SSP}$ for the sample are 0.35$\pm$0.09,
0.25$\pm$0.10 and 0.32$\pm$0.09~dex for V99, BC03 and TMB03 respectively.
We can again use the rms values to set an upper limit on the errors in 
[E/Fe]$_{SSP}$ of $\pm$0.1~dex, which is again consistent
with the rms deviations found in the test of robustness detailed in Section
\ref{fitting_proc}.

Interestingly, Fig. \ref{agez} suggests that values of [E/Fe]$_{SSP}$ in GCs 
are slightly higher than that of the local 
stars used to populate the SSPs \emph{at all metallicities}. The additional
enhancement in $\alpha$--elements, relative to that of the stellar 
calibrators, is 0.14, 0.05 and 0.11~dex in V99, BC03 and TMB03 SSP 
models respectively. It should be noted that these [E/Fe] excesses 
are independent of the assumed [E/Fe] pattern in the stellar calibrators, 
as the fitting technique outputs the [E/Fe] \emph{relative} to the models. 
While an enhancement with respect to the stellar calibrators is perhaps to be 
expected at high metallicities ([Fe/H]~$>$~--1), where the SSP models are based
largely on thin disc stars, the effect persists to the low metallicities at
which the models are constructed largely using thick disc and halo stars.
This is a slightly surprising result that clearly requires confirmation.

The stellar bulge data appear to indicate a relatively young age (6.3$\pm$3.6).
However, the P02 bulge data are highly susceptible
to stochastic fluctuations in the stellar contents of the fields from which
the spectra are constructed. Given this, and the large error on the age estimate,
no conclusion is drawn here regarding the age of the bulge. On the other hand, the average 
[Fe/H]$_{SSP}$ of -0.37~dex and [Z/H]$_{SSP}$ of -0.09~dex found for the bulge 
are consistent with the known distribution of bulge star metallicities (Ferreras, 
Wyse \& Silk 2003), while the [E/Fe]$_{SSP}$ of 0.29~dex is consistent with the 
$\alpha$--element enhancements from high resolution studies (e.g. McWilliam \& 
Rich 2004). Therefore, while no firm conclusion is reached with respect to age, the 
$\chi^2$--minimisation technique produces metallicity and abundance ratio 
estimates for the bulge which are in good agreement with literature values.

In summary, we find that the values of age, [Fe/H] and 
[E/Fe] derived using the $\chi^2$--minimisation technique are therefore in 
good agreement with values published in the literature and reproduce many of
the known properties of Galactic GCs and the bulge. Using the data of P02 and CBR98, upper 
limits on statistical errors in log(age), [Fe/H] and [E/Fe] have been shown to be $\pm$0.1~dex. 
Derived parameters with such errors could clearly provide powerful insights 
into the chemical and formation histories of extra-galactic GC systems and
their host galaxies.

\section{Conclusions}
We have compared the Lick index data of Puzia et al. (2002) and Cohen et al. 
(1998) for 20 Galactic GCs to the SSP models of Vazdekis (1999), Bruzual \&
Charlot (2003) and Thomas, Maraston \& Bender (2003). We
showed that ages derived from 2--dimensional index plots, such as
H$\beta$--[MgFe], are not reliable for these data. An alternative approach 
of using a $\chi^2$--minimisation technique was outlined. This technique allowed 
the identification of the known nitrogen over-abundance in Galactic GCs, as
well as interstellar absorption in NaD and poorly calibrated indices.
It was shown that this technique successfully 
recovers the known age, metallicity and abundance ratio properties 
of the Galactic GC population, reproducing values from the literature 
with an accuracy of $\sim$0.1~dex in any parameter.
However, even after poorly modelled indices were omitted from the
$\chi^2$--minimisation procedure, the fits to the models still 
exhibited reduced--$\chi^2$ values significantly 
in excess of 1.0 for all GCs in both studies. This implies that SSP 
modelling uncertainties are significant, and emphasises the advantages
of using a large number of indices in measuring derived parameters. 
We therefore conclude that the
$\chi^2$--minimisation technique holds great promise for the study of the
ages and metallicities of extra-galactic GC systems.\\

\noindent 

\noindent{\bf Acknowledgements}\\
We would like to thank Thomas Puzia and 
collaborators for providing data for this analysis. 
Our thanks also go to Jay Strader and Thomas Puzia for useful comments. 
The authors acknowledge the data analysis facilities provided by IRAF, which
is distributed by the National Optical Astronomy Observatories and
operated by AURA, Inc., under cooperative agreement with the National
Science Foundation. We thank the Royal Society and the Australian
Research council for funding that supported this work. \\

\noindent{\bf References}\\
\noindent
Alongi M., Bertelli G., Bressan A., Chiosi C., Fagotto F., Greggio L.,
  Nasi E., 1993, A\&AS, 97, 851\\
Beasley M.A., Hoyle F., Sharples R.M., 2002, MNRAS, 336, 168\\
Beasley M.A., Brodie J.P., Strader J., Forbes D.A, Proctor R.N., Barmby P.,
  Huchra J.P., 2004, astroph/0405009\\
Bensby T., Feltzing S., Lundstr{\o}m I., 2003, A\&A, 410, 527\\
Bertelli G., Bressan A., Chiosi C., Fagotto F., Nasi E., 1994, A\&AS, 106, 275\\
Bono G., Caputo F., Cassisi S., Castellani V., 1997, ApJ, 489, 822\\
Bressan A., Fagotto F., Bertelli G., Chiosi C., 1993, A\&AS, 100, 647\\
Bressan A., Chiosi C., Tantalo R., 1996, A\&A, 311, 425\\
Bruzual G., Barbuy B., Ortolani S., Bica E., Cuisinier F., Lejeune T.,
  Schiavon R.P., 1997, AJ, 114, 1531\\
Bruzual G., Charlot S., 2003, MNRAS, 344, 1000 (BC03)\\
Carney B.W., 1996, PASP, 108, 900\\
Carretta E., Cohen J.G., Gratton R.G., Behr B.B., 2001, AJ, 122, 1469\\
Cassisi S., Castellani M., Castellani V., 1997, A\&A, 317, 10\\
Cohen J.G., Blakeslee J.P., Ryzhov A., 1998, ApJ, 496, 808 (CBR98)\\
Cohen J.G., Gratton R.G., Behr B.B., Carretta E., 1999, ApJ, 523, 739\\
Edvardsson B., Andersen J., Gustafsson B., Lambert D.L., Nissen P.E.,
  Tomkin J., 1993, A\&A, 275, 101\\
Fagotto F., Bressan A., Bertelli G., Chiosi C., 1994a, A\&AS, 104, 365\\
Fagotto F., Bressan A., Bertelli G., Chiosi C., 1994b, A\&AS, 105, 29\\
Ferreras I., Wyse R.F. G., Silk J., 2003, MNRAS, 345, 1381\\
Girardi L., Bressan A., Chiosi C., Bertelli G., Nasi E., 1996, A\&AS, 117, 113\\
Gorgas J., Pedraz S., Guzman R., Cardiel N., Gonz\'{a}lez J.J., 1997, ApJ, 481, L19\\
Gustafsson B., Karlsson T., Olsson E., Edvardsson B., Ryde N., 1999, A\&A, 342, 426\\
Harris W.E., 1996, AJ, 112, 1487\\
Kuntschner H., Davies R.L., 1998, MNRAS, 295, L23\\
Kuntschner H., Smith R.J., Colless M., Davies R.L., Kaldare R., Vazdekis A.,
  2002, MNRAS, 337, 172\\
Li Y., Burstein D., 2003, ApJ, 598, L106\\
Longhetti M., Bressan A., Chiosi C., Rampazzo R., 2000, A\&A, 353, 917\\
  Maraston C., Greggio L., Renzini A., Ortolani S., Saglia R.P., Puzia
  T.H., Kissler-Patig M., 2003, A\&A, 400, 823\\
McWilliam A., Rich R.M., 2004, Origin and Evolution of the Elements,
  Carnegie Observatories Centennial Symposia. Carnegie Observatories
  Astrophysics Series. Ed. by A. McWilliam and M. Rauch,
  Carnegie Observatories, http://www.ociw.edu/ociw/symposia/series/symposium4/ 
  proceedings.html\\
Piotto G., King I.R., Djorgovski S.G., Sosin C., Zoccali M., Saviane I.,
  De Angeli F., Riello M., Recio Blanco A., Rich R.M., Meylan G., Renzini A.,
  2002, A\&A, 391, 945\\
Proctor R.N., 2002, Ph.D Thesis, University of Central Lancashire, Preston, UK \\ 
  (http://www.star.uclan.ac.uk/~rnp/research.htm)\\
Proctor R.N., Sansom A.E., 2002, MNRAS, 333, 517 (PS02)\\
Proctor R.N., Forbes, D.A., Hau G.K.T., Beasley, M.A., De Silva, G.M.,
  Contreras, R., Terlevich, A.I., 2004, MNRAS, 349, 1381\\
Puzia T.H., Saglia R.P., KIissler--Payig M., Maraston C., Greggio L., Renzini
  A., Ortolani S., 2002, A\&A, 395, 45 (P02)\\
Rich R.M., 1999, Spectrophotometric Dating of Stars and Galaxies, ASP Conference 
  Proceedings, Vol. 192, Edited by Ivan Hubeny, Sally Heap, and Robert Cornett, p.215\\
Rose J.A., 1984, AJ, 89, 1258\\
Rose J.A., 1985, AJ, 90, 1927\\
Rosenberg A., Saviane I., Piotto G., Aparicio A., 1999, AJ, 118, 2306\\
Saglia R.P., Maraston C., Thomas D., Bender R., Colless M., 2002, ApJ, 579,13\\
Salaris M., Weiss A., 2002, A\&A, 388, 492\\
Salasnich B., Girardi L., Weiss A., Chiosi C., 2000, A\&A, 361, 1023\\
Schiavon R.P., Faber S.M., Castilho B.V., Rose J.A., 2002a, ApJ, 580, 850\\
Schiavon R.P., Faber S.M., Rose J.A., Castilho B.V., 2002b, ApJ, 580, 873\\
Schiavon R.P., Rose J.A., Courteau S., MacArthur L.A., 2004, ApJ, 608, 33\\
Thomas D., Maraston C., 2003, A\&A, 401, 429\\
Thomas D., Maraston C., Bender R., 2003, MNRAS, 339, 897 (TMB03)\\
Thomas D., Maraston C., Korn A., 2004, MNRAS, 351, L19\\
Trager S.C., 2004, Origin and Evolution of the Elements, from the Carnegie
  Observatories Centennial Symposia. Published by Cambridge University Press,
  as part of the Carnegie Observatories Astrophysics Series. Edited by A.
  McWilliam and M. Rauch, p. 391.\\
Terlevich A.I., Forbes D.A., 2002, MNRAS, 330, 547\\
Trager S.C., Worthey G., Faber S.M., Bustein D., Gonz\'{a}lez J.J., 1998, ApJS, 116, 1\\ 
Trager S.C., Worthey G., Faber S.M., Gonz\'{a}lez J.J., 2000, AJ 119, 1645 (T00)\\
Tripicco M.J., Bell R.A., 1995, AJ, 110, 3035 (TB95)\\
Vazdekis A., 1999, ApJ, 513, 224 (V99)\\
Vazdekis A., Peletier R.F., Beckman J.E., Casuso E., 1997, ApJS, 111, 203\\
Worthey G., 1992, Ph.D. Thesis California Univ., Santa Cruz, USA\\
Worthey G., 1994, ApJS, 95, 107\\
Worthey G., Faber S.M., Gonz\'{a}lez J.J., 1992, ApJ, 398, 69\\
Worthey G., Ottaviani D.L., 1997, ApJS, 111, 377\\
\label{lastpage}
\end{document}